# Assessment Model for Opportunistic Routing

W. Moreira, P. Mendes and S. Sargento

*Abstract*— This paper proposes an assessment model, based on a new taxonomy, which comprises an evaluation guideline with performance metrics and experimental setup to aid designers in evaluating solutions through fair comparisons. Simulation results are provided based on the proposed model considering *Epidemic*, *PROPHET*, *Bubble Rap*, and *Spray and Wait*, and showing how they perform under the same set of metrics and scenario.

*Keywords*—Opportunistic Routing, Assessment Model.

## I. INTRODUÇÃO

A CAPACIDADE dos dispositivos portáteis permite aos utilizadores formar facilmente redes para partilha de recursos (i.e., processamento) e de conectividade. Tais redes espontâneas tiram vantagens dos contatos oportunistas entre os dispositivos dos utilizadores para armazenar e transmitir informação mesmo na presença de conectividade intermitente (resultante da mobilidade do utilizador, obstáculos físicos).

Existem várias propostas de encaminhamento que fazem uso da capacidade dos dispositivos para superar intermitência nas comunicações, onde os dispositivos armazenam dados até que um nó intermediário adequado, ou o próprio, destino seja encontrado com base no paradigma *store-carry-forward*. Estas propostas abrangem abordagens que consideram mobilidade do nó para inundar a rede com varias cópias das mensagens para uma entrega rápida, até abordagens capazes de controlar o número de cópias com base no histórico e previsão de contatos, por exemplo. Nos últimos anos surgiram abordagens baseadas na similaridade social (e.g., relações sociais, interesses) para melhorar o encaminhamento oportunístico.

Este artigo propõe um Modelo de Avaliação Universal (MAU) que tem um conjunto de parâmetros relacionados com a densidade e tráfego na rede, e é baseado numa classificação que identifica propriedades comuns (i.e., estratégias e métricas de encaminhamento) de entre propostas de encaminhamento oportunista, no sentido de ter um suporte no desenvolvimento de novas propostas. Também é comparado o desempenho de propostas oportunistas nas condições especificadas no MAU.

Este artigo está organizado da seguinte forma. A Seção II apresenta as classificações de estratégias de encaminhamento e modelos de avaliação existentes. A proposta de classificação e o modelo de avaliação propostos são apresentados nas Seções III e IV, respectivamente. Na Seção V é apresentada e discutida uma avaliação justa entre *Epidemic* [1], *PROPHET* [2], *Bubble Rap* [3], e *Spray and Wait* [4] baseado no modelo MAU. Finalmente, a Seção VI conclui o artigo.

## II. TRABALHOS RELACIONADOS

Propostas de encaminhamento oportunista não consideram um conjunto similar de métricas de desempenho nem cenários experimentais comparáveis. Isso resulta em comparações injustas entre propostas, uma vez que as métricas de avaliação e condições variam. Nesta seção apresentamos diferentes classificações de estratégias de encaminhamento e modelo de avaliação relacionados com a nossa proposta, MAU.

### A. Classificação das Estratégias de Encaminhamento

Existem diferentes classificações de estratégias de encaminhamento oportunista [5] que as categorizam: i) com base em oráculos de conhecimento da rede ou cálculo e determinação de rota [6]; ii) considerando se o comportamento da rede e dos nós é conhecido (i.e., determinístico) ou desconhecido e aleatório (i.e., estocástico) [7]; iii) levando em conta o conhecimento necessário sobre a rede e a estratégia de encaminhamento (i.e., inundação da rede, histórico de contatos e utilização de dispositivos especiais) [8]; e, iv) de acordo com o esquema de troca de mensagens (i.e., encaminhamento de uma única cópia, replicação e codificação), funções utilitárias e características (e.g, conectividade e mobilidade) que têm impacto no encaminhamento.

A tendência observada (i.e., encaminhamento baseado em similaridade social) nos últimos anos é mencionada, mas apenas como uma forma de utilizar dispositivos especiais [8] ou como mera função utilitária [9].

Verifica-se que o principal objetivo destas classificações é de exclusivamente identificar as diferentes famílias de soluções de encaminhamento. Algumas [6] [9] fornecem princípios de avaliação que apenas ajudam no desenvolvimento e identificação dos requisitos das aplicações, a fim de propor o algoritmo correto. Por outro lado, a nossa classificação identifica aspectos comuns entre as propostas analisadas para propor uma maneira justa de avaliar o seu desempenho, independentemente da quantidade de conhecimento necessário e cenário de aplicação.

### B. Modelos de Avaliação

Quanto aos modelos de avaliação, destaca-se uma proposta baseada na teoria de evolução dos grafos para desenvolver e avaliar protocolos com menor custo de encaminhamento. Ferreira et al. [10] usam uma métrica para determinar caminhos (i.e., ligações temporárias entre os nós que forneçam caminhos) que rapidamente chegam ao destino.

Este modelo fornece um algoritmo que é utilizado como referência para comparar propostas de encaminhamento em redes *adhoc* móveis. Contudo, não é discutido como as métricas de desempenho e parâmetros experimentais devem ser configuradas. Por outro lado, neste artigo pretende-se

W. Moreira, SITI, Universidade Lusófona, Lisboa, Portugal, waldir.junior@ulusofona.pt
P. Mendes, SITI, Universidade Lusófona, Lisboa, Portugal, paulo.mendes@ulusofona.pt
S. Sargento, IT, Universidade de Aveiro, Aveiro, Portugal, susana@ua.pt

identificar as diferentes famílias de propostas de acordo com os seus objetivos distintos e estratégias de encaminhamento, e para fornecer um conjunto de configurações experimentais para auxiliar na avaliação justa de soluções de encaminhamento oportunista.

## III. MODELO DE CLASSIFICAÇÃO

Observando-se as abordagens de encaminhamento oportunista presentes na literatura, é clara a existência de diferentes tendências com base em objetivos específicos. Assim, há a necessidade de uma taxonomia equilibrada e atualizada capaz de incluir as tendências identificadas nos últimos anos, com foco exclusivamente no ramo que melhor reflete contatos oportunistas (i.e., estocástico [7]). Esta seção descreve brevemente esta nova taxonomia (direciona-se o leitor para uma versão detalhada em [11]).

A taxonomia na Fig. 1 considera a análise de propostas anteriores e que evoluiu a partir de uma análise prévia feita em [12]. São identificadas três grandes categorias baseadas em encaminhamento de uma única cópia, inundação de mensagens e replicação. Entre as propostas consideradas, 94% pertencem à categoria de replicação incluindo a tendência identificada, similaridade social, que é de grande interesse para o nosso estudo e foco das discussões.

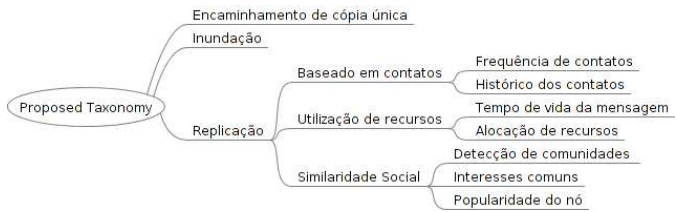

Figura 1. Taxonomia para Encaminhamento Oportunista.

A categoria de encaminhamento de cópia única é bastante interessante, já que estas propostas são capazes de poupar recursos da rede e do nó. No entanto, elas sofrem de uma elevada latência, o que resulta em baixa probabilidade de entrega. Exemplos desta categoria são apresentados em [13].

Embora sendo bastante agressivos, os algoritmos baseados em inundação de mensagens atingem elevada probabilidade de entrega, mas com custo elevado (i.e., consumo de recursos). Um exemplo desta categoria é o *Epidemic* [1].

A fim de reduzir o desperdício de recursos, propostas baseadas na replicação visam aumentar a probabilidade de entrega através da adição controlada de cópias de mensagem na rede. Estas propostas são divididas com base nos contatos, uso de recursos ou similaridade social.

Na categoria de propostas baseadas em contatos, é considerado o histórico de contatos com o destino para dar suporte ao encaminhamento oportunista de mensagens. A frequência de contatos com nós no passado, ou o tempo decorrido desde o último contato com o destino é usado para decidir quando encaminhar (e.g., *PROPHET*).

As propostas baseadas na utilização de recursos tentam evitar que as mensagens sejam constantemente replicadas (i.e., ocupando recursos) através de métricas que definem tempo de vida para as cópias da mensagem. Estas propostas também são capazes de tomar decisões de encaminhamento que utilizam eficientemente os recursos disponíveis (e.g., *RAPID* [14]).

Na categoria de similaridade social, as propostas têm algoritmos mais complexos que visam: i) evitar a inundação descontrolada de mensagens e, ii) explorar o comportamento social relacionado com a detecção da comunidade, interesses comuns, e a popularidade do nó. O *Bubble Rap* [3] é um bom representante desta categoria.

## IV. MODELO DE AVALIAÇÃO UNIVERSAL

O estudo proposto neste artigo mostra que os métodos de avaliação utilizados até agora nem sempre consideram um conjunto homogêneo de parâmetros ou configurações experimentais comparáveis, o que compromete a veracidade e imparcialidade das conclusões. Mesmo com uma taxonomia estável, há a necessidade de conceber um modelo de avaliação com base em métricas de desempenho e cenários experimentais comuns, evitando avaliações com as métricas de desempenho irrelevantes e cenários tendenciosos.

Assim, estudamos dezessete propostas (de 2000 a 2010) que melhor representam as categorias da nossa taxonomia para identificar padrões de avaliação/comparação. Mais detalhes deste estudo encontram-se no documento [11].

### A. Métricas de Desempenho

Num estudo anterior [12] observou a falta de uma convenção de nomenclatura e diversas definições para uma determinada métrica. Este entendimento diferente certamente influencia a avaliação de desempenho. Observa-se também que a probabilidade de entrega, custo e latência são as métricas mais utilizadas. Sendo assim, elas são definidas para estabelecer uma convenção de nomenclatura.

*Probabilidade de entrega* é definida como o número de mensagens transmitidas em relação ao número total de mensagens criadas. Esta métrica é importante, pois reflete a eficácia proposta classificando-a como tendo bom desempenho. No entanto, há sempre um *custo* associado para a entrega das mensagens no que diz respeito ao número de cópias por mensagem entregue. Além disso, a utilidade da mensagem é correlacionada com o seu tempo de vida (TTL), e é imperativo para remover dados antigos da rede a fim de evitar o desperdício de recursos. Por isso, é importante que as mensagens cheguem ao seu destino dentro do tempo útil. Assim, define-se *latência* como o tempo necessário para entregar uma mensagem desde sua criação.

### B. Cenário experimental

Outro ponto importante é a configuração experimental, onde não há regras definidas já que os parâmetros variam nas avaliações encontradas na literatura.

Pudemos observar que as propostas realmente diferem quanto ao cenário experimental. Algumas propostas fornecem informações detalhadas, enquanto outras as fornecem apenas parcialmente.

TABELA I
PARÂMETROS DE DENSIDADE DA REDE E DE TRÁFEGO.

| Parâmetros | (a) Densidade da rede | | | | | (b) Tráfego | | | | |
|---|---|---|---|---|---|---|---|---|---|---|
| Propostas | Densidade da área (Km²/Número de nós) | Modelo de Mobilidade Velocidade (m/s) | Tempos de contato/ Inter-contato | Alcance de Transmissão (m) | Emissão de beacon | Distribuição de Fontes/Destinos | Número de Mensagens Geradas (Carga) | Tamanho da Mensagem (Kb) | TTL da Mensagem | Tamanho do *Buffer* |
| MAU | Cobrindo áreas dentro de prédios até cidades/100 a 150 | Representando pessoas e diferentes meios de transporte | Exponencial ou lei de potência, se puder ser determinado | Mínimo de 10 (Bluetooth) e máximo de 250 (Wi-Fi) | A cada 100 ms | Deve ser o mesmo durante toda a avaliação | Carga deve ser a mesma | 1 a 100 | 3 a 10 saltos ou variação entre horas, dias e semanas | 2MB (200 msgs de 10KB) |
| PROPHET | 0.45/50 (RWP), e 4.5/50 (CM) | Mobilidade *Random Waypoint* (0-20) e baseado em Comunidade (10-30) | Nós entram em contato periódica e randomicamente | 50 e 100 | n/a | 45/44 (Random) e 2/1 (Com.) | 1 mensagem/seg (RWP), e 20 msgs/seg – 2msgs/5seg (com.) | n/a | 3 e 11 saltos | 200 msgs |
| Bubble Rap | Infocom05/41 Hong-Kong/37 Cambridge/54 Infocomo6/98 Reality/97 | *Traces* reais | Tempo de inter-contato segue uma distribuição de lei de potência | Bluetooth | n/a | 1/1 | 1000 mensagens | n/a | 2, 10 min, 1, 3, 6 horas 1, 2, 4 dias 1, 3 semanas | n/a |

Também pudemos identificar duas classes principais de parâmetros: de densidade de rede e de tráfego. As Tabelas I-a e I-b mostram o que recomenda o MAU, juntamente com o que é usado no *PROPHET* e *Bubble Rap*. Para mais informações sobre *Epidemic*, *Spray and Wait* e outras propostas, recomenda-se a leitura do documento [11].

A densidade da rede permite entender o funcionamento das propostas em cenários esparsos (contatos esporádicos, observando-se a probabilidade de entrega quando a latência é alta) e densos (contatos frequentes, avaliando a capacidade de lidar com a aleatoriedade na hora do encaminhamento). A densidade pode ser ajustada através do número de nós, que dentre as propostas analisadas, fica entre 100-150 (excluindo casos extremos [15]), e são os valores considerados no MAU.

Além disso, os modelos de mobilidade devem considerar diferentes velocidades e tempo de pausa já que os nós representam pessoas e veículos. Estes parâmetros têm impacto nos tempos de contato e intercontato. Observamos que algumas abordagens consideram estes tempos como distribuições exponenciais e de lei de potência [14] [3], enquanto outras os obtêm dos conjuntos de dados utilizados.

Analisamos também o alcance de transmissão, onde propomos variar entre 10-250 metros, uma vez que se deve representar as capacidades de dispositivos. Quanto aos *beacons*, deve-se evitar o uso frequente já que a bateria precisa ser poupada. No entanto, usando-os raramente resulta em perda de boas oportunidades de contato. Assim, a emissão de *beacons* a cada 100 ms pode identificar potenciais contatos, enquanto economiza energia. Ressaltamos que este parâmetro no MAU requer um estudo mais detalhado para ser validado.

Em relação aos parâmetros de densidade de rede (cf. Tabela I-a), o *PROPHET* considera o mesmo número de nós (50) para diferentes densidades, modelos de mobilidade (incluindo um modelo que tenta imitar o comportamento humano) e considera diferentes alcances de transmissão.

Apesar de considerar diferentes densidades, modelos de mobilidade e alcances de transmissão, a avaliação do *PROPHET* é feita em cenários homogêneos. Isto não é realista, pois o cenário real é heterogêneo onde há nós que se deslocam de acordo com padrões e velocidade diferente e os dispositivos têm capacidades de transmissão específicas.

Quanto ao *Bubble Rap*, que tem sua avaliação baseada em *traces* humanos, os autores conseguem ter densidades diferentes (áreas cobrindo salas de conferências até uma cidade). Embora não mencione sobre modelos de mobilidade, o *Bubble Rap* segue o comportamento humano encontrado nos *traces*, o que também resulta em tempos de contato e intercontato adequados [14] [3]. O alcance de transmissão considerado foi o do *Bluetooth*, que representa os dispositivos utilizados na obtenção dos *traces*.

Já o *Spray and Wait* tem a sua avaliação considerando diferentes alcances de transmissão e variando o número de nós, contudo utiliza apenas um modelo de mobilidade que está longe do encontrado no mundo real.

Quando comparado ao *PROPHET* e *Spray and Wait*, o *Bubble Rap* destaca-se em termos de parâmetros aceitáveis referente à densidade da rede proposta no MAU. No entanto, a sua avaliação é feita de forma estática, ou seja, as comunidades são formadas e as centralidades são determinadas com base em informações que não são atualizadas. O nosso interesse é ver como a proposta se comporta num cenário dinâmico, que influencia a forma com que a centralidade e as comunidades são calculadas.

Quanto aos parâmetros de tráfego (cf. Tabela I-b), em quase todas as propostas, o número de pares fonte-destino foi definido estaticamente e estes foram distribuídos aleatoriamente. Não há problema em ter este número definido estaticamente, uma vez que mudá-lo durante as experiências vai impor alguns desafios. No entanto, este número deve ser o mesmo e, mais importante, os pares devem permanecer os mesmos (o que não é assegurado com a distribuição aleatória) para garantir condições semelhantes no estudo de avaliação.

A carga gerada é um parâmetro que acrescenta mais variações às experiências. Observamos propostas onde a carga corresponde a uma mensagem por segundo [1] [2], um número de mensagens distribuídas uniformemente [3], bem como propostas que fornecem pouca ou nenhuma informação sobre a mesma. Acreditamos que a carga deva ser considerada cuidadosamente para garantir comparações justas.

Como os aplicativos geram mensagens de diferentes tamanhos, a carga deve ser ajustada para representar as aplicações existentes num cenário oportunista (e.g., mensagens de chat, transferência de arquivos). Observamos que poucas propostas fornecem informações sobre este parâmetro (1 KB [1] [14] ou entre 10-100 KB [16]).

O tráfego na rede pode ser afetado pelo TTL da mensagem. Se o TTL é muito longo, o consumo de recursos na rede e nos nós pode aumentar. Caso contrário, as mensagens talvez nem alcancem o destino. Nas propostas estudadas o TTL foi definido como o número de saltos [1] [2] ou unidades de tempo [4], e, normalmente varia entre 3-10 saltos em média. Hui et al. [3] mostra que apenas 5% dos nós tem algum nível de relação com o destino no primeiro salto, assim, sugerimos pelo menos 3 saltos (já que os níveis de interação entre os nós melhora em cerca de 35%) com variação para observar o comportamento das propostas. No caso de se utilizar unidades de tempo, pudemos observar que um TTL em 24 horas e também infinito pode mostrar potenciais problemas (e.g., rápida exaustão de *buffer*) com a proposta sendo avaliada.

Observamos, também, como o desempenho da proposta pode ser influenciado pelo tamanho do *buffer*. Este parâmetro reflete o quanto do armazenamento do seu dispositivo um utilizador está disposto a sacrificar pelos outros. O *buffer* ilimitado não é realista, enquanto que fornecer todo o espaço é aceitável em cenários onde nós (e.g., ônibus) devem servir aos outros [14]. Assim, com base nas nossas observações das propostas, deve-se limitar o *buffer* em 200 mensagens (10KB).

Da Tabela I-b, observamos que o *PROPHET* tem uma configuração diferente, onde a carga varia de acordo com o modelo de mobilidade, e são considerados diferentes valores de TTL. Normalmente, o *buffer* pode ser um problema, já que os utilizadores podem não estar dispostos a compartilhá-lo. Assim, considerar um *buffer* limitado está mais perto do cenário encontrado no mundo real, e é uma opção dos autores do *PROPHET*.

No entanto, em relação a estes parâmetros, *Bubble Rap* só fornece informações sobre os pares fonte-destino e de carga (1000 mensagens geradas entre todos os pares de nós e de tempo de vida das mensagens). O mesmo ocorre com o *Spray and Wait*, onde as mensagens são geradas seguindo uma distribuição uniforme para destinos escolhidos aleatoriamente.

Os parâmetros de tráfego são aqueles que mais variam nos estudos de comparação. Normalmente, os pares fonte-destino são escolhidos aleatoriamente. Isto influencia a avaliação, já que este conjunto vai variar conforme as simulações são realizadas. Assim, os estudos de avaliação devem considerar o mesmo subconjunto de pares fonte-destino, e o número de mensagens geradas também deve permanecer o mesmo.

Nestas propostas, os autores não mencionam sobre o tamanho da mensagem. Este parâmetro é importante visto que as aplicações geram mensagens com tamanhos diferentes. Apesar de tentar poupar o *buffer*, o *Bubble Rap* não indica o tamanho das mensagens que foi considerado na sua avaliação.

O *Bubble Rap*, assim como o *Spray and Wait*, considera o TTL (expressos em número de saltos) das mensagens e sugere um valor mínimo não inferior a 3, já que a maioria dos nós encontrados nos primeiros saltos ainda pertencem à comunidade do nó que transporta a mensagem. No entanto, o TTL é representado em unidades de tempo na sua avaliação.

## V. AVALIAÇÃO JUSTA

Nesta seção apresentamos a avaliação de desempenho do *Epidemic*, *PROPHET*, *Bubble Rap* e *Spray and Wait* nas mesmas condições especificadas no MAU. É importante notar que pretendemos mostrar a importância de uma avaliação homogênea ao avaliar propostas de encaminhamento de categorias diferentes como as baseadas em inundação, contatos e similaridade social, e mostrar como o uso de diferentes configurações de parâmetros e métricas de desempenho pode resultar numa avaliação que favoreça algumas propostas.

### A. Configurações de Avaliação e Metodologia

O cenário escolhido tem 150 nós distribuídos em 8 grupos de pessoas e 9 grupos de veículos, onde um destes grupos representa patrulhas policiais e segue o modelo de mobilidade *Shortest Path Map Based Movement* [17] com velocidade entre 7-10 m/s e tempos de pausa entre 100-300 segundos.

Os outros grupos de veículos representam ônibus, onde cada grupo tem 2 veículos, e seguem o modelo de mobilidade *Bus Movement* [17] com velocidades entre 7-10 m/s e tempos de pausa entre 10-30 segundos.

Os grupos de pessoas seguem o modelo de mobilidade *Working DayMovement* [17] com velocidades entre 0.8-1.4 m/s, mas podem usar os ônibus para se deslocar. Cada um destes grupos tem diferentes pontos de encontro, escritórios e moradas. As pessoas passam 8 horas por dia no trabalho e têm 50% de probabilidade de terem uma atividade após o trabalho. No escritório, elas movimentam-se e têm tempos de pausa entre 1 minuto e 4 horas. As atividades após o trabalho podem ser feitas a sós ou em grupos e duram entre 1 a 2 horas.

Cada nó tem uma interface de rede sem fios com velocidade de transmissão de 11 Mbps e alcance de 100 metros.

O tráfego é estabelecido entre os mesmos pares fonte-destino, onde cerca de 500 mensagens são geradas por dia. O tamanho das mensagens varia de 1 a 100 kB e o TTL varia de horas a semanas. O *buffer* é de 2 MB. Estes valores representam as aplicações diferentes e a disposição limitada do utilizador em compartilhar o seu espaço de armazenamento.

As simulações são realizadas num *Opportunistic Network Environment* [17] e representam uma interação de 12 dias entre nós (com 2 dias de *warm-up*, que não são considerados para os resultados). Cada simulação é executada dez vezes (com diferentes sementes geradoras de números aleatórios para os modelos de mobilidade utilizados) a fim de proporcionar um intervalo de confiança de 95% para os resultados. Todos os resultados são avaliados considerando as médias da probabilidade de entrega, do custo e da latência.

### B. Resultados

Neste artigo observamos o comportamento das propostas sob as mesmas condições e cenários utilizados em [5], contudo, agora dando mais ênfase ao TTL das mensagens, visto que há diferentes aplicações em redes oportunistas que geram mensagens com vida útil variada.

A Fig. 2 apresenta os resultados obtidos para a probabilidade média de entrega. Ao comparar *Epidemic* e *PROPHET* no cenário definido através do MAU, observa-se o mesmo comportamento relatado por Lindgren et al. [2]: *PROPHET* tem melhor desempenho, alcançando uma vantagem que chega até 17 pontos percentuais.

Entretanto, para um TTL de uma hora, o *Epidemic* tem mais vantagem por inundar a rede rapidamente com cópias das mensagens, o que aumenta a sua probabilidade de entrega. Porém, a proposta sofre com o aumento do TTL pois as mensagens passam mais tempo no sistema (i,e., a replicar-se) e, consequentemente, reduz a sua probabilidade de entrega, já que o *buffer* atinge o seu limite mais rapido.

tem menor custo. Tomar decisões de encaminhamento com base em aspectos sociais acaba por diminuir o custo já que as cópias apenas serão criadas entre nós bem relacionados entre si. Contudo, não considerar o dinamismo dos relacionamentos sociais afeta negativamente este tipo de abordagem [18].

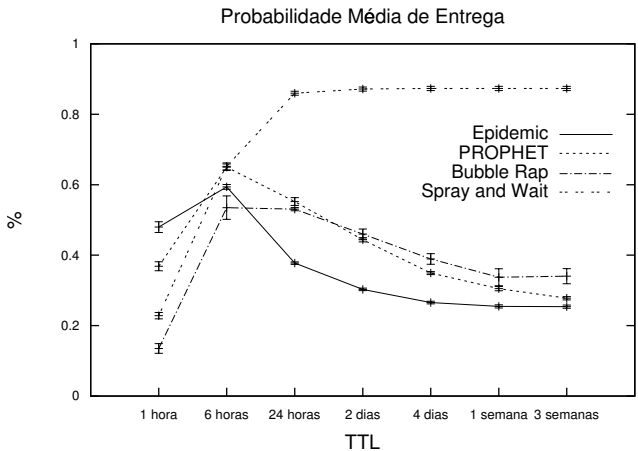
Figura 2. Resultados de Probabilidade Média de Entrega.

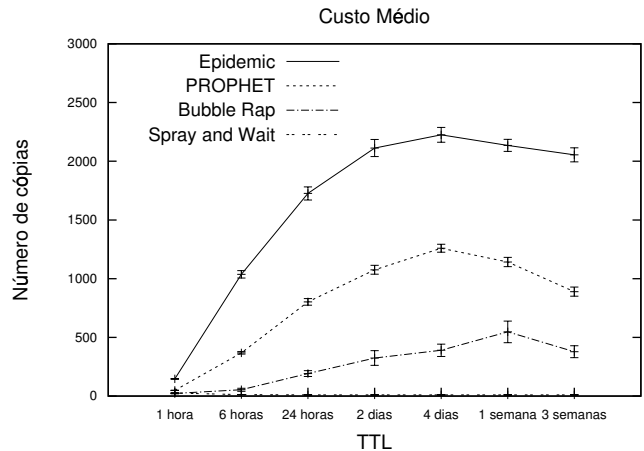
Figura 3. Resultados de Custo Médio.

No que concerne à comparação de desempenho entre o *PROPHET* e o *Bubble Rap* no cenário especificado com base no MAU, observamos três pontos interessantes ao contrário do que é reportado no artigo original do *Bubble Rap*. Primeiro, a probabilidade de entrega começa a diminuir para os TTLs acima das 6 horas. Observa-se também que o *Bubble Rap* apresenta melhores taxas de entrega que o *PROPHET* para TTL maior ou igual a 2 dias. Além disso, ambas propostas chegam próximo aos 60% de entrega apenas quando o TTL é de 3 semanas, enquanto que no nosso cenário o *PROPHET* ultrapassa (~65%) e o *Bubble Rap* se aproxima (~54%) desta marca com um TTL de 6 horas. Considera-se que este é o efeito não só do *buffer* limitado, mas também da heterogeneidade no que concerne os modelos de mobilidade utilizados no cenário.

O *Spray and Wait* é aquele que tem a melhor probabilidade de entrega no cenário. Isto justifica-se pelo fato de o cenário conter nós (i.e., ônibus e patrulhas de polícia) que o cobrem quase que em toda a sua extensão. Como estes nós acabam por se deslocar muito no cenário, aliado também ao alcance de transmissão (100 m), consequentemente aumenta-se muito a probabilidade das mensagens chegarem ao destino.

A Fig. 3 apresenta os resultados relacionados com o custo médio das propostas. Pode-se dizer que o *Epidemic* e *Spray and Wait* são os casos extremos no que concerne esta métrica. O primeiro funciona com base na inundação descontrolada (i.e., cada nó recebe uma cópia se ainda nao a tiver) e assim o faz para ter uma taxa elevada de entrega. O problema é que o nível de interação entre os nós aliado a um *buffer* limitado torna-se um grande inimigo deste tipo de abordagem. Em contra-partida, o *Spray and Wait* tem o número de cópias (L = 10) por mensagem limitado e o seu custo vai ser sempre baixo.

Quanto ao *PROPHET* e *Bubble Rap*, a diferença entre eles permanece como no artigo original do *Bubble Rap* onde este

O *PROPHET* teve menor custo quando comparado com o *Epidemic* como reportado no seu artigo original. Entretanto, o seu custo é muito maior (~34 vezes) para um TTL de 24 horas, por exemplo.

A Fig. 4 apresenta a latência média para mensagens entregues de cada proposta. Tanto o *Epidemic* como o *Spray and Wait* têm um comportamento similar no que concerne esta métrica. O primeiro, ao inundar a rede com cópias, acaba por alcançar nós que logo estarão em contato com o destino. Além disso, também contribui para o desempenho do *Spray and Wait* juntamente com o alcance de transmissão que, ao aumentar, diminui o tempo para as mensagens chegarem ao destino como reportado no seu artigo original.

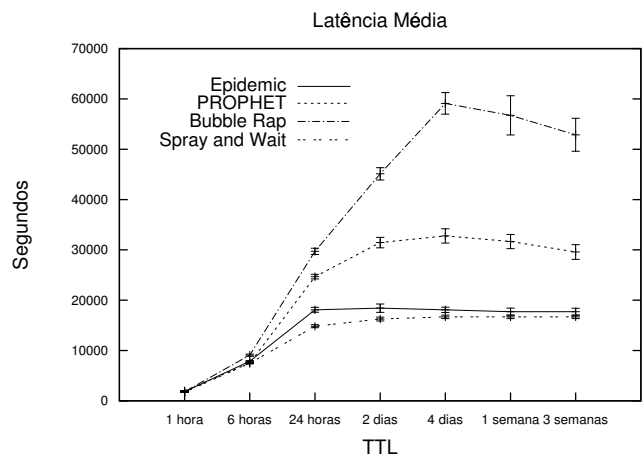
Figura 4. Resultados de Latência Média.

O *PROPHET* e o *Bubble* Rap, por serem propostas mais elaboradas, demoram mais tempo para decidir quando encaminhar mensagens. Para além disso, no caso do *Bubble Rap*, o fato de não considerar o dinamismo dos relacionamentos sociais faz com que as mensagens sejam encaminhadas a nós que demoram a encontrar o destino, o que contribui para o aumento da sua latência.

## VI. Conclusões

A capacidade atual dos dispositivos portáteis permite aos utilizadores experimentar novas formas de troca de informação por meio de contatos oportunistas. No entanto, esta nova forma de comunicação tem que lidar com a intermitência destes contatos, o que deu origem a diferentes propostas que tentam diminuir este problema. Contudo, é difícil identificar qual proposta apresenta melhor desempenho uma vez que cada uma delas tem um método de avaliação diferente.

Analisando as propostas dos últimos dez anos, observaram-se diferentes tendências com base em objetivos específicos e com diferentes soluções de encaminhamento oportunistas. Assim, neste artigo, analisamos propostas diferentes de acordo com as tendências identificadas, as informações colecionadas no seu processo de avaliação, e descobrimos propriedades comuns entre as várias propostas (ou seja, estratégias e métricas de encaminhamento). O resultado deste estudo é um Modelo de Avaliação Universal (MAU) que fornece diretrizes, baseadas numa taxonomia, incluindo a nova tendência identificada recentemente (i.e., a similaridade social), compreendendo um conjunto de parâmetros de desempenho e configuração experimental para auxiliar a avaliação do desempenho de soluções de encaminhamento oportunistas. Para validar os nossos princípios, simulamos as propostas *Epidemic*, *PROPHET*, *Bubble Rap*, e *Spray and Wait* sob as mesmas condições e pudemos verificar que as diferenças entre as propostas são mais evidentes no MAU.

Acreditamos que atingimos o nosso objetivo de fornecer uma maneira para efetuar a classificação das soluções novas de encaminhamento oportunista, bem como avaliá-las adequadamente. Como surgem novas propostas na literatura, esta taxonomia será atualizada com as últimas tendências identificadas no contexto das redes oportunistas.

### Agradecimentos

A FCT pelo apoio financeiro a Waldir Moreira (PhD grant SFRH/BD/62761/2009) e ao projeto UCR (PTDC/EEA-TEL/103637/2008).

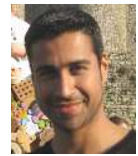
**Waldir Moreira** é graduado (05) em Ciência da Computação pela Universidade da Amazônia, e obteve seu grau de mestre (08) em Ciência da Computação pela Universidade Federal do Pará, em Belém, Brasil. Atualmente, é aluno do MAP Doctoral Programme in Telecommunications na Universidade de Aveiro e é membro do Internet Architectures and Networking Lab na unidade de investigação em Sistemas e Tecnologias Informáticas da Universidade Lusófona. Seus interesses de pesquisa são encaminhamento oportunista e baseado em aspectos sociais em redes mesh, cooperativas, e tolerantes a atrasos/disrupções.

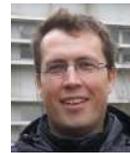
**Paulo Mendes** é graduado (93) em Engenharia Informática pela Universidade de Coimbra, Mestre (98) em Engenharia Electrotécnica e de Computadores pela Universidade Técnica de Lisboa, Ph.D. (03) em Engenharia Informática pela Universidade de Coimbra. Atualmente, é diretor científico para a inovação da unidade de investigação em Sistemas e Tecnologias Informáticas da Universidade Lusófona, coordenador do Internet Architectures e Networking Lab, e Professor Associado da Universidade Lusófona. É membro da ACM SIGCOMM e contribui para o IETF/IRTF. Os seus interesses científicos são redes cooperativas, redes auto-organizativas sem fios e sistemas sensoriais cooperativos. Paulo Mendes tem mais de 60 artigos e 13 patentes.

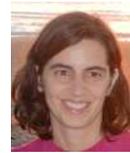
**Susana Sargento** é graduada (97) em Engenharia Electrónica e de Telelcomunicações pela Universidade de Aveiro, Ph.D. (03) em Engenharia Electrotécnica pela Universidade de Aveiro. Atualmente, Susana foi docente no Departamento de Ciências de Computadores da Universidade do Porto de 2002 a 2004, e encontra-se desde 2004 na Universidade de Aveiro e Instituto de Telecomunicações. Durante os últimos anos ela tem estado envolvida em vários projetos nacionais e europeus, com responsabilidades de coordenação de várias atividades, como exemplo FP6 IST-Daidalos. Está correntemente envolvida em vários projetos Europeus FP7 (Euro-NF), nacionais, e da CMU|Portugal (DRIVE-IN com a Universidade de Carnegie Melon). É membro do IEEE. Os seus interesses de investigação centram-se nas áreas de redes heterogêneas, infra-estruturadas, em malha e ad-hoc (foco em veicular).